\documentclass[aps,prb,twocolumn,showpacs,a4paper,floatfix]{revtex4-1}
\usepackage{times}
\usepackage[latin1]{inputenc}
\usepackage[english]{babel}
\usepackage[T1]{fontenc}
 \usepackage{graphicx}
\usepackage{epstopdf}
\usepackage[small,FIGTOPCAP]{subfigure}
\usepackage{varioref}
\usepackage{rotating}
\usepackage{mathptmx}
\usepackage{fancyhdr}
\usepackage{amsmath}
\usepackage{amsfonts}
\usepackage{amssymb}
\usepackage{amsbsy}
\usepackage{textcomp}
\usepackage{array}
\usepackage{fixmath}
\usepackage{bm}
\usepackage[active]{srcltx}
\usepackage[colorlinks]{hyperref}
\usepackage[all]{hypcap}

\hypersetup{ pdfcreator=Emiliano Cadelano}

\makeindex             % used for the subject index
\begin{document}
\title{The effect of the hydrogen coverage on the Young modulus of graphene}
\author{Emiliano Cadelano$^1$ and Luciano Colombo$^{1,2}$}
\email{luciano.colombo@dsf.unica.it}
\affiliation{$^1$Istituto Officina dei Materiali del Consiglio Nazionale delle Ricerche (CNR-IOM), Unit\`a di Cagliari,  Cittadella Universitaria I-09042 Monserrato (Ca), Italy}
% \author{Luciano Colombo}
\affiliation{$^2$Department of Physics, University of Cagliari, Cittadella Universitaria, I-09042 Monserrato (Ca), Italy }

% \affiliation{Istituto Officina dei Materiali del Consiglio Nazionale delle Ricerche (CNR-IOM), Unit\`a di Cagliari,  Cittadella Universitaria, I-09042 Monserrato (Ca), Italy}

\begin{abstract}
We blend together continuum elasticity and first principles calculations to measure by a computer experiment the Young modulus of hydrogenated graphene. 
We provide evidence that hydrogenation generally leads to a much smaller longitudinal extension upon loading than in pristine graphene.
Furthermore, the Young modulus is found to depend upon the loading direction for some specific conformers, characterized by an anisotropic linear elastic behavior.
\end{abstract}

\maketitle

The hydrogenated form of graphene\cite{Sluiter, sofo, boukhvalov, elias} (also referred to as graphane) has been systematically investigated  by Wen {\it et al.} \cite{wen}, proving that in fact there exist eight graphane isomers. They all correspond to covalently bonded hydrocarbons with a C:H ratio of $1$. Interesting enough, four isomers have been found to be more stable than benzene, indeed an intriguing issue.
By hydrogenation, 
a set of two dimensional materials is generated with new physico-chemical properties: for instance, 
graphane is an insulator\cite{sofo,boukhvalov}, with an energy gap as large as $\sim 6$ eV \cite{lebegueGW}, while graphene is a highly conductive semi-metal. 
Furthermore, disordered hydrogenated samples 
show electronic and phonon properties quite different than the pristine material\cite{elias}.

As far as the elastic behavior is concerned, it has been proved that hydrogenation largely affects the elastic moduli as well.  
By blending together continuum elasticity theory and first principles calculations, Cadelano \textit{et al.} \cite{cadelano2} have determined the linear and non linear elastic moduli of three stable graphane isomers, namely : chair- (C-), boat- (B-), and washboard- (W-) graphane. The resulting picture is very interesting: B-graphene is found to have a small and negative Poisson ratio, while,  due to the lack of isotropy, C-graphane admits both softening and hardening non linear hyperelasticity, depending on the direction of applied load.

Although full hydrogen coverage is possible both in ordered and disordered configurations,\cite{wen} it is more likely that a typical experimental processing procedure generates samples with a C:H  ratio larger than $1$. In other words, we must admit that graphane could exist not only in a large variety of conformers, included the amorphous one (hereafter referred to as A-graphane), but also in several forms characterized by different stoichiometry.

In this work we investigate the variation of the Young modulus of graphane versus the hydrogen coverage, in three stable conformers \cite{cadelano2} and in the amorphous one. The work is addressed to establish whether an incomplete $sp^{3}$ hybridization affects the elastic behavior and which is the trend (if any) of variation of the Young modulus versus hybridization, an issue of large technological impact.
Our approach combines continuum elasticity (used to define the deformation protocol aimed at determining the elastic energy density of the investigated systems) and first principles atomistic calculations (used to actually calculate such an energy density and the corresponding elastic moduli).

Atomistic calculations have been performed by Density Functional Theory (DFT) as implemented in the \textsc{Quantum ESPRESSO} package \cite{quantum-espresso}. 
The exchange correlation potential was evaluated through the generalized gradient approximation (GGA) with the Perdew-Burke-Ernzerhof (PBE) parameterization \cite{PBE}, using Rabe-Rappe-Kaxiras-Joannopoulos (RRKJ) ultrasoft pseudopotentials \cite{RRKJ, marzari}. A plane wave basis set with kinetic energy cutoff as high as 24 Ry was used and the Brillouin zone (BZ) has been sampled by means of a (4x4x1) Monkhorst-Pack grid. The atomic positions of the investigated samples have been optimized by using damped dynamics and periodically-repeated simulation cells. Accordingly, the interactions between adjacent atomic sheets in the supercell geometry were hindered by a large spacing greater than 10 \AA. 

The Young modulus of the structures under consideration has been obtained from energy-vs-strain curves, corresponding to suitable deformations applied to samples with different hydrogen coverage, namely: 25\%, 50\%, 75\%, and 100\%. 
The corresponding simulation cell
contained 8 carbon atoms and 2, 4, 6, and 8 hydrogen atoms, respectively.
For any possible coverage (other than the 100\% one), several different geometries (up to 10) have been considered, by randomly distributing hydrogen atoms according to different decoration motifs. This implies that all data below are obtained through configurational averages, a technical issue standing for the robustness of the present results.

For any deformation the magnitude of the strain is represented by a single parameter $\zeta$.\cite{cadelano2} Thus, the energy-vs-strain curves have been carefully generated by varying the magnitude of $\zeta$ in steps of  0.001 up to a maximum strain $\zeta_{max}=\pm 0.02$. This choice warrants that the linear elastic regime was carefully explored. 
All results have been confirmed by checking the stability of the estimated elastic moduli over several fitting ranges for each sample.
The reliability of the present computational set up is proved by the estimated values for the Young modulus ($E$) and the Poisson ratio ($\nu$) of pristine graphene (corresponding to 0\% of hydrogen coverage), respectively 349 Nm$^{-1}$ and 0.15, which are in excellent agreement with recent literature.\cite{cadelano2,kudin,gui,liu,cadelano} Similarly, our results $E=219$ Nm$^{-1}$ and $\nu=0.21$ for stoichiometric C-graphane (corresponding to 100\% hydrogen coverage) agree with data reported in Ref. \onlinecite{cadelano2, topsakal, munoz}.

\begin{figure}[tbp]
\begin{center}
\includegraphics[width= 0.48\textwidth, angle=0]{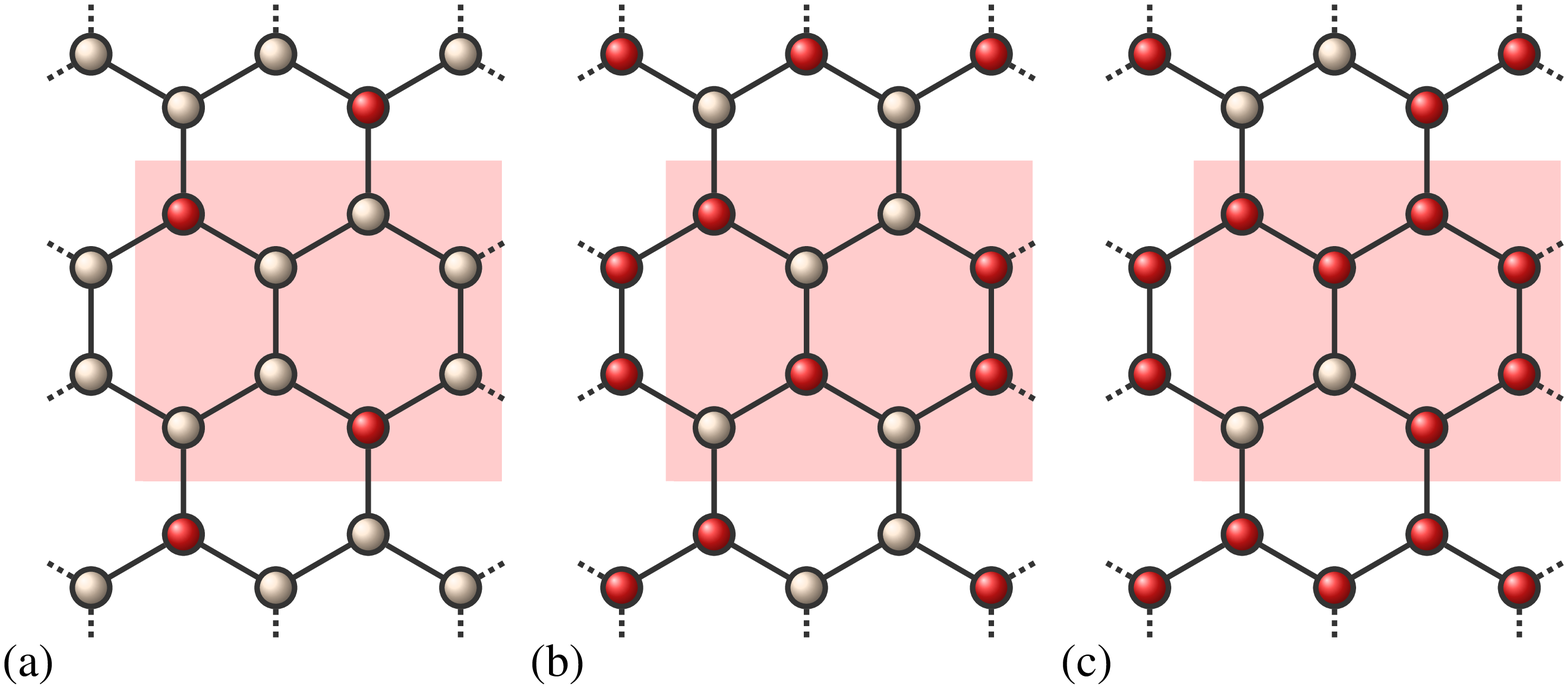}
\caption{\label{gra2}
Pictorial representations of some hydrogen motifs corresponding to 25\% (Panel a), 50\% (Panel b), and 75\% (Panel c) coverage. Hydrogen atoms are indicated by red (dark) circles, while hydrogen vacancies by gray (light) circles which are randomly placed on the top or bottom of the graphene sheet (black sticks). Shaded areas represent the simulation cell. Up to 10 different decorations for each conformer have been used in the calculations.
}
\end{center}
\end{figure}

Among the systems here investigated, stoichiometric C-graphane is elastically isotropic due to its crystallographic symmetry,
while its non stoichiometric conformers with 25\%, 50\% and 75\% hydrogen coverage (see Fig. \ref{gra2}) are so by assumption (which is indeed reasonable by only assuming that the hydrogen decoration in real samples is totally random).
Similar properties hold for A-graphane. On the other hand, stoichiometric B- and W-graphane show an orthorhombic symmetry, which causes an anisotropic linear elastic behavior. Their non stoichiometric conformers could became isotropic due to the achieved disorder degree, as discussed below.

The two-dimensional elastic energy density (per unit of area) for systems with orthorhombic symmetry (namely, B- and W-graphane) is expressed as \cite{huntington, landau}
\begin{eqnarray}
U= \frac{1}{2}\mathcal{C}_{11}\epsilon_{xx}^{2}+\frac{1}{2}\mathcal{C}_{22}\epsilon_{yy}^{2}+\mathcal{C}_{12} \epsilon_{xx}\epsilon_{yy}+2\mathcal{C}_{44}\epsilon_{xy}^{2}
\label{orto}
\end{eqnarray}
where  $x$ ($y$)  indicates the zigzag (armchair) direction in the honeycomb carbon lattice and $\epsilon_{ij}$ ($i,j=x~or~y$) is the infinitesimal strain tensor.
By applying a loading at an angle $\theta$ with respect to the zigzag direction, the corresponding Young modulus $E(\theta)$ is written as\cite{cadelano2}
\begin{equation}
\label{EEE}
E(\theta)= \frac{\mathcal{C}_{11}\mathcal{C}_{22}-\mathcal{C}_{12}^2}{\mathcal{C}_{11}s^4+\mathcal{C}_{22}c^4 +\left( \frac{\mathcal{C}_{11}\mathcal{C}_{22}-\mathcal{C}_{12}^2}{\mathcal{C}_{44}}-2 {\mathcal{C}_{12}}\right)c^2 s^2}
\end{equation}
where  $ c=\cos \theta$, and $ s=\sin \theta $.
Eqs.(\ref{orto}) and (\ref{EEE}) are also valid for isotropy systems (namely, C- and A-graphane) provided that it is set: 
$\mathcal{C}_{11}=\mathcal{C}_{22} $ and  $ 2\mathcal{C}_{44}=\mathcal{C}_{11}-\mathcal{C}_{12} $ (Cauchy relation).

The Young modulus $E(\theta)$ can be directly obtained from the linear elastic constants $\mathcal{C}_{ij}$, in turn computed through energy-vs-strain curves corresponding to suitable homogeneous in-plane deformations.
Only two in-plane deformations must be applied to obtain all the independent elastic constants, namely: (i) an uniaxial deformation along the zigzag (or armchair) direction; and (ii) an hydrostatic planar deformation. Nevertheless, for the validation of the isotropicity condition, two more in-plane deformations must be further applied: (iii) an axial deformation along the armchair (or zigzag) direction; and (iv) a shear deformation. 
The strain tensors corresponding to applied deformations depend on the unique scalar strain parameter $ \zeta $ \cite{cadelano2, cadelano}, so that the elastic energy of strained structures defined in Eq.(\ref{orto}) can be written as
$U(\zeta)=U_0+\tfrac{1}{2}U^{(2)}\zeta^2+O(\zeta^3)$,
where $U_0$ is the energy of the unstrained configuration. Since the expansion coefficient $U^{(2)}$ is related to the elastic constants $\mathcal{C}_{ij}$, a straightforward fit provides the full set of linear moduli for all structures.
Repeating this procedure for several randomly distributed hydrogens in different configurational symmetries (corresponding to C-, B-, W-, and A-graphane), the complete set of elastic constants $\mathcal{C}_{ij}$ has been evaluated (as an average value) at various coverages.

\begin{table}[bp]
\caption{\label{tablefin} Independent elastic constants (units of  Nm$^{-1}$) are shown for different values of the hydrogen coverage. The 0\% coverage corresponds to the pristine graphene, where $\mathcal{C}_{11}$=357 $\pm$7, and $\mathcal{C}_{12}$ 52$\pm$11.  }
\begin{tabular}{c r l r l  r l r l}
\hline
\hline
\             &\ \  \ \  \ \ 25\% \ \ & &\ \  \ \  \ \ 50\% \ &\ &\ \  \ \  \ \ 75\%&\ &\  \ \  \ \ 100\% \ & \\
\hline
\multicolumn{9}{c}{(C-graphane)} \\
$\mathcal{C}_{11}$        & \  267 $\pm$& 8      & \ 227 $\pm$&12        & \ 258 $\pm$&7       & \ 230 $\pm$&10 \\
$\mathcal{C}_{12}$       & \  51 $\pm$& 16   &  \  17 $\pm$&27   & \  10 $\pm$&11  & \ 50 $\pm$&20 \\
\multicolumn{9}{c}{(A-graphane)} \\
$\mathcal{C}_{11}$         & \   267 $\pm$& 4  & \ 225 $\pm$&5       & \   194$\pm$&7       & \  190 $\pm$&9 \\
$\mathcal{C}_{12}$        & \  32 $\pm$& 7 &  \  19 $\pm$&8   & \  12 $\pm$&10   & \  16$\pm$&8 \\
\multicolumn{9}{c}{(B-graphane)} \\
$\mathcal{C}_{11}$        & \  274 $\pm$& 3      & \ 243 $\pm$&11      & \ 214.0 $\pm$&6       & \ 259 $\pm$&12 \\
$\mathcal{C}_{22}$      & \ 254 $\pm$& 5   &  \  209 $\pm$& 8 & \  164 $\pm$&5  & \ 228 $\pm$&11 \\
$\mathcal{C}_{12}$       & \  24 $\pm$& 6      & \  29 $\pm$&14           & \ 5 $\pm$&7       & \ -2 $\pm$&4 \\
$\mathcal{C}_{44}$       & \ 121 $\pm$& 2   &  \  98 $\pm$&4   & \  133 $\pm$& 2 & \ 94 $\pm$&10 \\
\multicolumn{9}{c}{(W-graphane)} \\
$\mathcal{C}_{11}$       & \  274 $\pm$& 6      & \ 243 $\pm$&8        & \ 250 $\pm$&6       & \ 303 $\pm$&5 \\
$\mathcal{C}_{22}$       & \ 258 $\pm$& 8   &  \  186 $\pm$&11   & \  71 $\pm$&10  & \ 75 $\pm$&4 \\
$\mathcal{C}_{12}$        & \  32 $\pm$& 10      & \ 7 $\pm$&10        & \ 22 $\pm$&8       & \ 14 $\pm$&6 \\
$\mathcal{C}_{44}$      & \ 119 $\pm$& 4   &  \  101 $\pm$&5   & \  65 $\pm$&9  & \ 69 $\pm$&2 \\
\hline
\hline
\end{tabular}
\end{table}

The synopsis of the calculated elastic constants $\mathcal{C}_{ij}$ for all graphane samples here investigated is reported in Table \ref{tablefin}, from which quite a few information can be extracted. As a preliminary remark, we stress that each hydrogenated conformer is characterized by a specific hydrogen arrangement and by a different buckling of the carbon sublattice. 
Moreover, in non stoichiometric samples hydrogen atoms can migrate (as indeed observed during relaxation) due to the presence of nearby unsaturated bonds, thus affecting the resulting symmetry of the conformer.
These features add further details to an already complex situation, introducing another source of disorder. We, therefore, argue that the amorphous configuration should be considered the most probable phase.  
Alternatively, the above mentioned migration of hydrogen atoms could indeed generate phase separation into C-H and pure C domains, as discussed by Lin \textit{et al.} \cite{liu2}.

\begin{figure}[tbp]
\begin{center}
\includegraphics[width= 0.48\textwidth, angle=0]{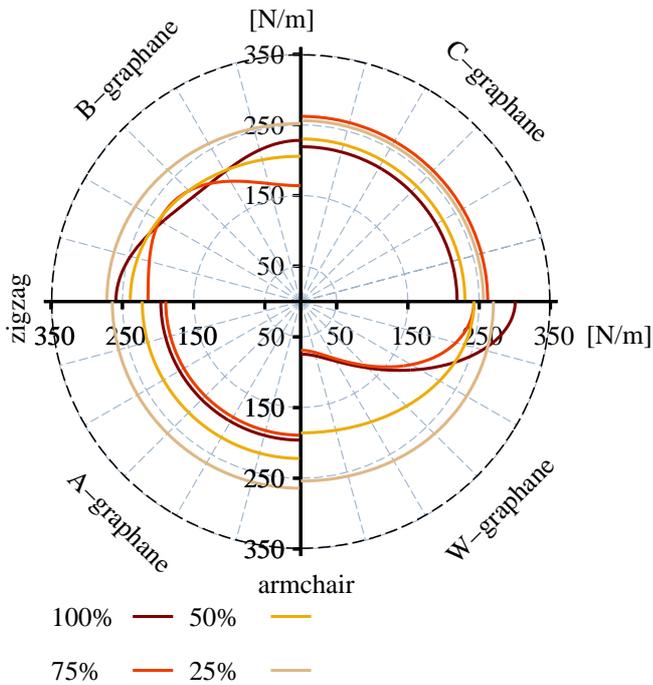}
\caption{\label{gra1}
Polar plot of the Young modulus $E(\theta)$ (see Eq.(\ref{EEE})) shown as function of the hydrogen coverage. In each quadrant a given conformer of graphane is represented, as indicated by labels. The Young modulus of pristine graphene is also shown by the bold dashed line.}
\end{center}
\end{figure}

By inserting the values reported in Table \ref{tablefin} into the Eq.(\ref{EEE}), we obtain the Young modulus $E(\theta)$ for any hydrogen coverage, as shown in Fig.\ref{gra1}.
As a general feature,  we state that the change in hybridization has largely reduced the property of longitudinal resistance upon extension, as described by the greatly reduced value of the Young modulus with respect to ideal graphene (bold dashed line in Fig. \ref{gra1}). Simulations provide evidence that this is mainly due to the fact that $sp^{3}$ hybridization creates tetrahedral angles (involving 4 carbons and 1 hydrogen) which are easily distorted upon loading. In other words, softer tetrahedral deformations are observed, rather than bond stretching ones as in ideal graphene.
Equivalently, we can conclude that deformations upon loading are basically accommodated by variations of the tetrahedral angles.

The full picture about the elastic response is provided in Fig.\ref{gra1}, by considering the $\theta$-dependence of the calculated Young modulus for each graphane conformer. The corresponding value of ideal graphene is reported as well for comparison. In such a polar plot, an isotropic behavior corresponds to a circular line. This is always found for C- and A-graphane (as well as for ideal graphene), as expected. 
On the other hand, above 75\% hydrogen coverage, the B- and W-conformers show a strong $\theta$-dependence in their elastic behavior (or, equivalently, of their Young modulus). Interesting enough, well below this coverage they almost behave isotropically. 

Deviations from the isotropic elastic behavior are quantitatively predicted by the calculation of the $\mathcal{A}=4\mathcal{C}_{44}/(\mathcal{C}_{11}+\mathcal{C}_{22}-2\mathcal{C}_{12})$ ratio, which should be $1$ for ideally isotropic systems. Data reported in Table \ref{tablefin} provide  $\mathcal{A}=~1.0 \pm 0.05$ for any C- and the A-conformers, thus confirming their elastic isotropicity. On the other hand, non stoichiometric B- and W-samples with hydrogen coverage smaller than 75\% display a comparatively small anisotropy, corresponding to $\mathcal{A}=~1.0 \pm 0.2$. Finally, these same systems with higher hydrogen coverage recover a full anisotropic behavior ($\mathcal{A}$ is largely different than 1).

In conclusion, we have presented and discussed first principles calculations showing that the elastic behavior of graphene is largely affected by hydrogen absorption by the actual coverage. In particular, the Young modulus is greatly reduced upon hydrogenation, as also previously discussed in Ref. \onlinecite{munoz}. An incomplete coverage  generates a large configurational disorder in the hydrogen sublattice, leading to a larger corrugation with respect to highly-symmetric C-graphane. Indeed, such a corrugation of the carbon sublattice is a key feature affecting the overall elastic behavior.

We acknowledge financial support by Regional Government of Sardinia under the project "Ricerca di Base" titled "Modellizzazione Multiscala della Meccanica dei Materiali Complessi" (RAS-M4C). We acknowledge computational support by CYBERSAR (Cagliari, Italy) and CASPUR (Roma, Italy) computing centers.

\end{document}